*A revised version*

# First-principles calculations of a high-pressure synthesized compound PtC

**Linyan Li[1], Wen Yu[1,2], and Changqing Jin[1,*]**

[1] Institute of Physics, Chinese Academy of Sciences, PO Box 603, Beijing 100080, People's Republic of China

[2] Physics Department, University of Science and Technology Beijing, Beijing 100080, People's Republic of China

[*] Author to whom correspondence should be addressed

E-mail: **cqjin@aphy.iphy.ac.cn**

**Abstract**

First-principles density-functional method is used to study the recently high-pressure synthesized compound PtC. It is confirmed by our calculations that the platinum carbide has a zinc-blende ground-state phase at zero pressure and the rock-salt structure is a high-pressure phase. The theoretical transition pressure from zinc-blende to rock-salt is determined to be 52GPa. Furthermore, our calculation shows the possibility that the experimentally synthesized PtC by Ono *et al.* under high pressure condition might undergo a transition from rock-salt structure to zinc-blende after the pressure quench to ambient condition.

## 1. Introduction

Some carbides, especially transition-metal carbides, are widely used for industrial application because of their remarkable physical properties. In a recent paper, Shigeaki Ono *et al.* [1] synthesized a new platinum carbide. The compound is formed at high temperature and high pressure and was found to be stable after quenching to room temperature and ambient pressure. They thought the new PtC has a cubic symmetry and a rock-salt type structure with space group Fm3m by analyzing the synchrotron x-ray diffraction pattern. But it is impossible to distinguish between the zinc-blende (space-group number 216) and rock-salt (space-group number 225) structures solely from the x-ray diffraction because of the large mass difference between Pt and C. In this case, the Raman spectrum is an important criterion. For example, B.Gregoryanz *et al.* [2] synthesized PtN at high temperature and high pressure and the crystal structure of PtN has a cubic symmetry from the X-ray diffraction. They concluded that the PtN has a zinc-blende structure because the rock-salt structure doesn't have a first-order Raman spectrum. In this paper, we investigated the relative stability of the two different structures for PtC and discussed the possible phase transitions under pressure from the theoretical point of view.

## 2. Method of calculation

We use first-principles method to calculate the zinc-blende and rock-salt structures stability of PtC as a function of pressure. All calculations were performed using the Vienna package WIEN2k [4]. This is a full-potential linearized augmented plane waves (LAPW) within the density-functional theory (DFT) [5]. This method is one of the most accurate schemes in solving the Kohn-Sham equations. The generalized gradient approximation (GGA) [6] was used for the exchange and correlation potential function. In the zinc-blende structure the muffin-tin (MT) sphere radii of 2.2 and 1.5 bohr were used for Pt and C atoms respectively, while in the rock-salt structure 2.3 and 1.7 bohr were used because the lattice constants of the two structures are different. The valence wave functions inside the MT spheres were expanded into spherical harmonics up to $l=10$ and the $R_{mt}K_{max}$ were taken to be 8.0. We used 3000 k points in the Brillouin zone for the zinc-blende structure and 4000 k points for the rock-salt structure. The self-consistent calculations were considered to be converged only when the integrated magnitude of the charge density difference between input and output [i.e., $\int \left| \rho_{n+1}(\vec{r}) - \rho_n(\vec{r}) \right| d\vec{r}$] was less than 0.00001.

## 3. Results and discussion

### 3.1. Structural properties

The zinc-blende and rock-salt structures are shown in Figure 1. In the zinc-blende structure, the Pt atoms form a face-centered cubic lattice and the C atoms occupy half of the tetrahedral interstitial sites of the Pt lattice. In the rock-salt structure, the two face-centered cubic lattices formed respectively by Pt and C atoms are interpenetrated. Figure 2 is a plot of total energy as a function of unit-cell volume for the two structures. It shows that the zinc-blende structure for PtC has the lower energy minimum and that the minimum occurs at a larger lattice constant than for rock-salt structure. This means that the zinc-blende structure is the ground-state phase at zero pressure and at a sufficiently high pressure the rock-salt structure would be favored.

Since all our calculations are performed at $T = 0$ K, then the Gibbs free energy, $G = E_0 + PV - TS$, becomes equal to the enthalpy, $H = E_0 + PV$. Therefore, the phase transition pressure is determined with the tangent to the curves in figure 2. The slope of the tangent line reveals that, at a pressure of about 52 GPa, a phase transition from zinc-blende to rock-salt would occur if the temperature were high enough to break the zinc-blende covalent bonds.

The lattice constants, bulk modulus, and total energies of two structures for PtC are listed in Table 1. The bulk modulus ($B$) was evaluated from the Murnaghan equation fitting of the total energies as a function of the unit cell volume. The theoretical zinc-blende lattice constant is in agreement with experiment while the rock-salt lattice constant is 7% smaller than the experimental value of the quenched PtC sample. The total energy per formula unit of the zinc-blende structure is about 1.17 eV smaller than that of the rock-salt structure.

*3.2. Elastic properties*

In order to further confirm the structure stability under strain, we calculated the elastic constants of the two structures, which were shown in table 2. There are three independent elastic constants, e $c_{11}$, $c_{12}$, *and* $c_{44}$ for a cubic crystal. A stable cubic crystal should match with the conditions: $c_{44}>0$, $c_{11}> |c_{12}|$, and $c_{11}+2c_{12}>0$. The three elastic constants can be obtained from the second-order derivative of the total energy of the crystal under three type of shear strains: the volume change, the volume conserved tetragonal and rhomohedral strains. The bulk modulus evaluated from the elastic constants (Table 2) were consistent with those obtained from Murnaghan equation (Table 1) for the two structures of PtC. The rock-salt bulk modulus we calculated from the Murnaghan equation is 265 GPa which is about 12% smaller than the experimental value 301GPa. In the rock-salt structure, $c_{11} < |c_{12}|$ and in the zinc-blende structure, the restrictions are all fulfilled. With these results, we conclude that the zinc-blende structure is the stable ground state and the rock-salt structure is not stable at zero pressure and there is a possibility that a phase transition from rock-salt (high-pressure phase) to zinc-blende (ground-state phase) would occur after the pressure quench to ambient condition.

*3.3. Electronic properties*

Figure 3 displays the total and partial densities of states (DOS) for the zinc-blende PtC. The rock-salt DOS is similar to that of zinc-blende. No energy gap is seen, indicating a metallic nature of platinum carbide. At the Fermi level the total DOS is 0.88 states/eV unit cell. The DOS of Pt is 0.41 states/eV unit cell and that of C is 0.21. It is evident that the main DOS is from the Pt (Pt 5d) contribution. The states between -14 and -11 eV are mainly composed of C (2s) states. The states above –8eV are mainly composed of Pt (5d) states and C (2p) states. Due to the stronger Pt ($t_{2g}$)-C (2p) interactions, the $t_{2g}$ band has a larger dispersion than that of the $e_g$ band.

Figure 4 displays the energy bands and the characteristic bands along high-symmetry directions in the Brillouin zone. The band around -12 eV originates almost completely from C 2s states. The bands around -3eV and -4eV originate mainly from Pt $e_g$ states. From -7 eV to 1 eV, the bands are hybridization between Pt $t_{2g}$ and C 2p. From the characteristic bands of Pt $e_g$, Pt $t_{2g}$ and C 2p, the major contribution to the covalent bonds comes from the hybridization between Pt 5d and C 2p .

## 4. Conclusions

In summary, we have studied PtC in zinc-blende and rock-salt structures using first-principles calculations. With all the calculations we conclude that the zinc-blende structure of PtC is the ground-state phase at zero pressure while the rock-salt structure is a high-pressure phase. At a pressure of about 52 GPa, a phase transition from zinc-blende to rock-salt may occur. Our calculation also shows the possibility that the experimentally synthesized PtC by Ono *et al.* under high pressure condition might undergo a transition from rock-salt structure to zinc-blende after the pressure quench to ambient condition.

**Acknowledgement**

This work was partially supported by the national nature science foundation of China, the state key fundamental research project (2002CB613301).

**Figure captions:**

Figure 1. Unit cell schematics of (a) the zinc-blende structure and (b) the rock-salt structure. The large and small spheres denote Pt and C atoms, respectively.

Figure 2. Total energy as a function of the volume per formula unit for PtC in zinc-blende (solid line) and the rock-salt (dashed line) structures.

Figure 3. Total and partial densities of state of PtC in zinc-blende structure.

Figure 4. Energy bands and characteristic bands of zinc-blende PtC along high-symmetry directions near the Fermi enengy (located at $E$=0).

Tables:

Table 1. Lattice constants, bulk modulus, and total energy with respect to the PtC in the zinc-blende and rock-salt structures.

|         | Exp.  | zinc-blende(cal.) | rock-salt(cal.) |
|---------|-------|-------------------|-----------------|
| a (Å)   | 4.814 | 4.737             | 4.494           |
| B (Gpa) | 301   | 230               | 257             |
| E (eV)  |       | -1.17             | 0.0             |

Table 2. Elastic constants of PtC in the zinc-blende and rock-salt structures.

|             | $c_{11}$(Gpa) | $c_{12}$(Gpa) | $c_{44}$(Gpa) | $B=(c_{11}+2c_{12})/3$(Gpa) | $c_{11}-|c_{12}|$ |
|-------------|---------------|---------------|---------------|------------------------------|-------------------|
| zinc-blende | 284           | 199           | 70            | 227                          | >0                |
| rock-salt   | 252           | 271           | 55            | 265                          | <0                |

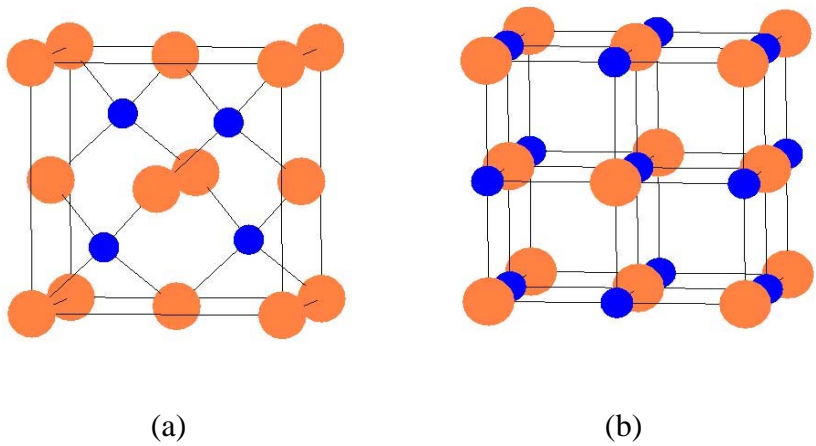

(a)          (b)

Figure 1.

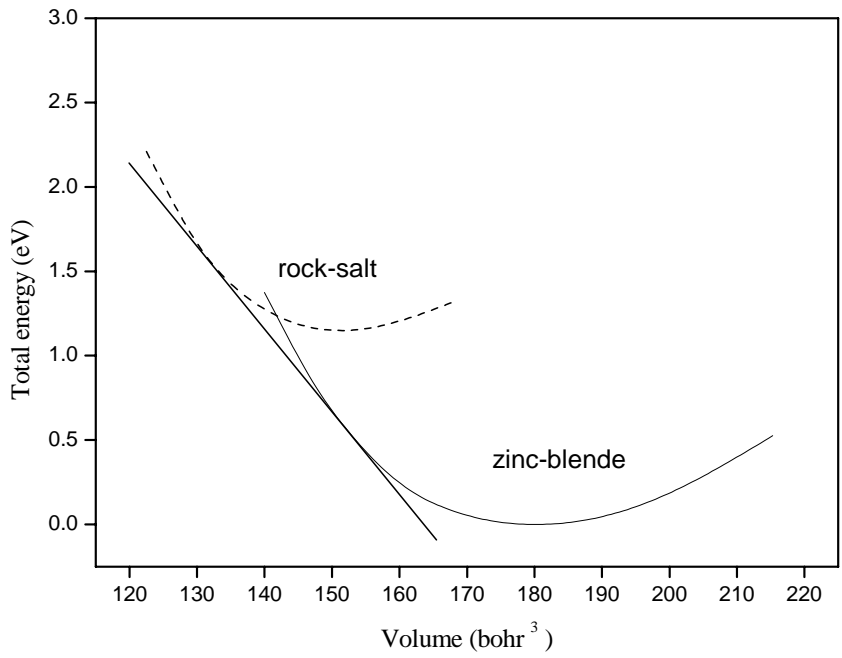

Figure 2.

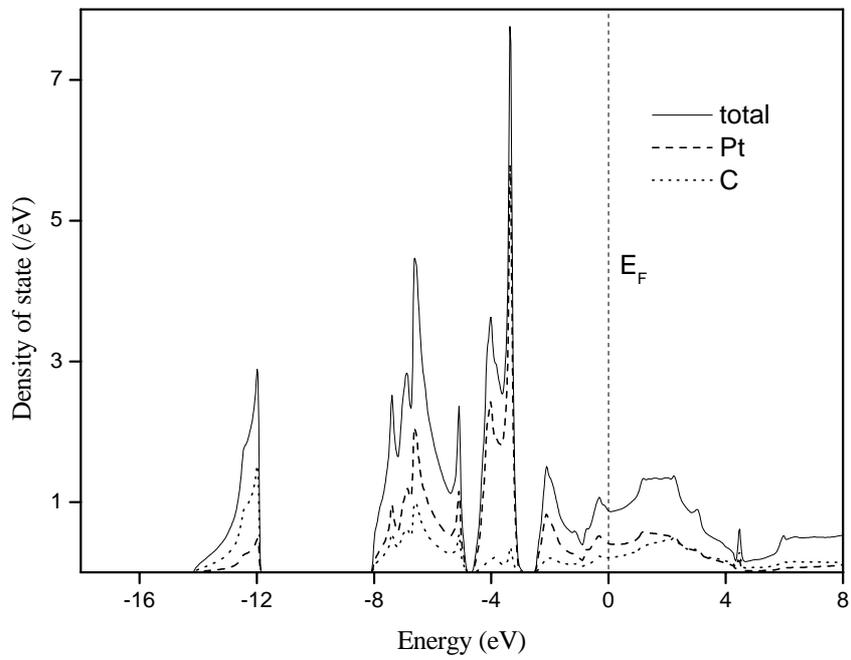

Figure 3.

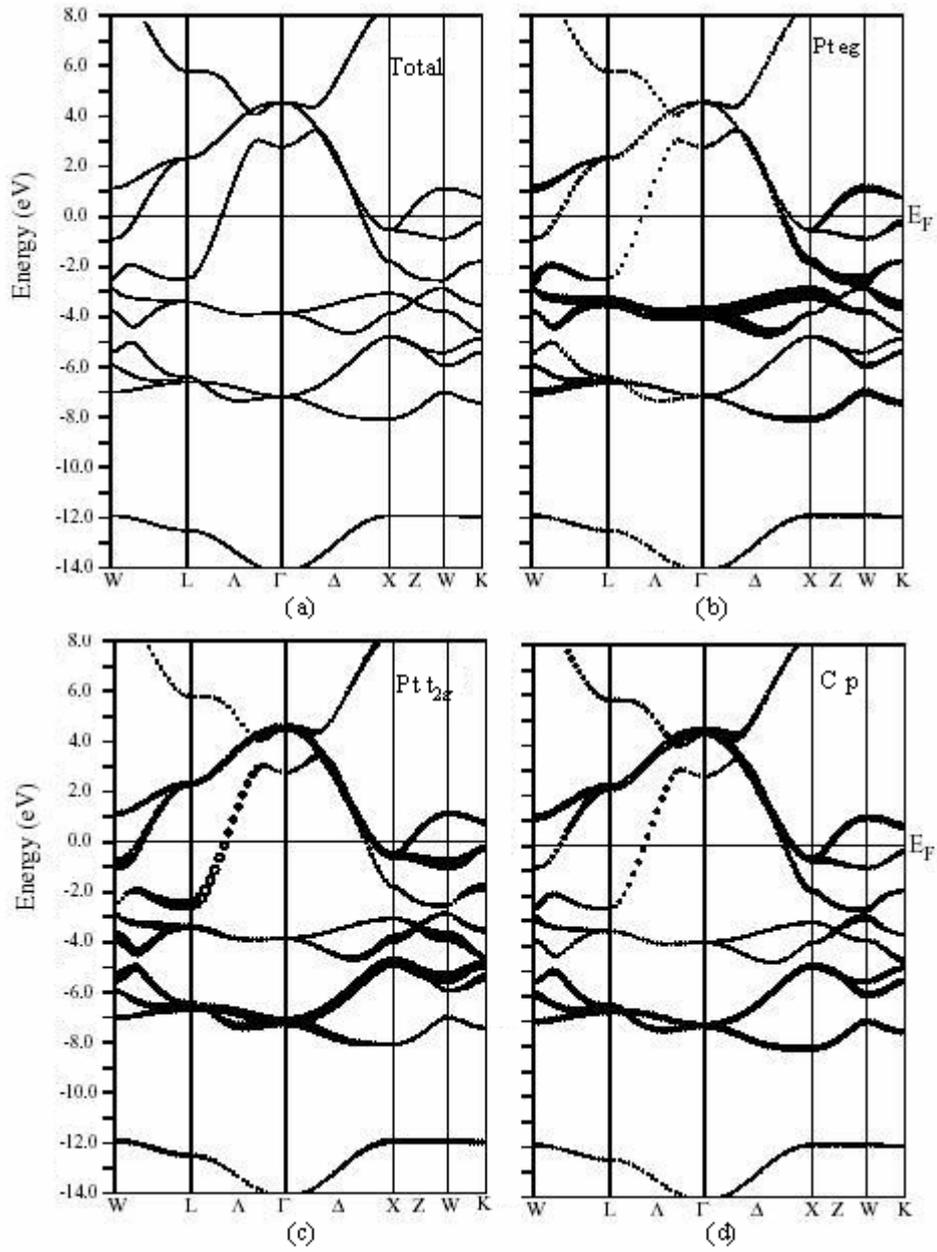

Figure 4.